\documentclass[%
 reprint,
 superscriptaddress,
 amsmath,amssymb,
 aps,
 prl,
 footinbib,
]{revtex4-1}

\usepackage{graphicx}
\usepackage{color}
\usepackage{dcolumn}
\usepackage{bm}

\begin{document}

\title{Magnetic and structural quantum phase transitions in CeCu$_{6-x}$Au$_x$ are independent}

\author{K. Grube}
 \email{kai.grube@kit.edu}
\author{L. Pintschovius}
\author{F. Weber}
\affiliation{Institut f\"ur Festk\"orperphysik, Karlsruher Institut f\"ur Technologie, D-76021 Karlsruhe, Germany}
\author{J.-P. Castellan}
\affiliation{Institut f\"ur Festk\"orperphysik, Karlsruher Institut f\"ur Technologie, D-76021 Karlsruhe, Germany}
\affiliation{Laboratoire L\'eon Brillouin (CEA-CNRS), CEA-Saclay, F-91911 Gif-sur-Yvette, France}
\author{S. Zaum}
\author{S. Kuntz}
\author{P. Schweiss}
\affiliation{Institut f\"ur Festk\"orperphysik, Karlsruher Institut f\"ur Technologie, D-76021 Karlsruhe, Germany}
\author{O. Stockert}
\affiliation{Max-Planck-Institut f\"ur Chemische Physik fester Stoffe, D-01187 Dresden, Germany}
\author{S. Bachus}
\author{Y. Shimura}
\author{V. Fritsch}
\affiliation{Experimentalphysik VI, Elektronische Korrelationen und Magnetismus, Universit\"at Augsburg, D-86159 Augsburg, Germany}
\author{H. v. L\"ohneysen}
 \email{hilbert.loehneysen@kit.edu}
\affiliation{Institut f\"ur Festk\"orperphysik, Karlsruher Institut f\"ur Technologie, D-76021 Karlsruhe, Germany}
\affiliation{Physikalisches Institut, Karlsruher Institut f\"ur Technologie, D-76049 Karlsruhe, Germany}

\date{\today}

\begin{abstract}
The heavy-fermion compound CeCu$_{6-x}$Au$_x$ has become a model system for unconventional magnetic quantum criticality. For small Au concentrations $0 \leq x < 0.16$, the compound undergoes a structural transition from orthorhombic to monoclinic crystal symmetry at a temperature $T_{s}$ with $T_{s} \rightarrow 0$ for $x \approx 0.15$. Antiferromagnetic order sets in close to $x \approx 0.1$. To shed light on the interplay between quantum critical magnetic and structural fluctuations we performed neutron-scattering and thermodynamic measurements on samples with $0 \leq x\leq 0.3$. The resulting phase diagram shows that the antiferromagnetic and monoclinic phase coexist in a tiny Au concentration range between $x\approx 0.1$ and $0.15$. The application of hydrostatic and chemical pressure allows to clearly separate the transitions from each other and to explore a possible effect of the structural transition on the magnetic quantum critical behavior. Our measurements demonstrate that at low temperatures the unconventional quantum criticality exclusively arises from magnetic fluctuations and is not affected by the monoclinic distortion.
\end{abstract}

\maketitle


Competing interactions often lead to magnetic quantum critical points (QCPs), where quantum fluctuations govern how a system develops long-range magnetic order. In metals, these fluctuations have a crucial impact on the electronic properties and usually lead to strong deviations from Fermi-liquid behavior. Heavy-fermion (HF) compounds, i.e., intermetallic compounds containing a sublattice of 4$f$ or $5f$ elements, have become model systems to study QCPs because small pressures suffice to shift the delicate balance between the Ruderman-Kittel-Kasuya-Yoshida (RKKY) interaction promoting long-range magnetic order and the Kondo effect suppressing ordering by locally screening the magnetic moments \cite{Doniach1977}. Many antiferromagnetic quantum-critical HF systems can be described within the Hertz-Millis-Moriya theory of QCPs, but there is a growing number of compounds that fall outside this classification \cite{HvL2007}, notably the HF compounds CeCu$_{6-x}$Au$_x$ \cite{HvL1999} and YbRh$_2$Si$_2$ \cite{Gegenwart2008}. 
In the model of local quantum criticality of HF systems, it was suggested that this unconventional behavior arises from the coincidence of different zero-temperature transitions: the onset of long-range antiferromagnetic order and the breakdown of the Kondo screening \cite{QSi2001}. More detailed studies revealed that in CeCu$_{6-x}$Au$_x$ the unconventional magnetic QCP is close to a structural transition from an orthorhombic ($Pnma$) to a monoclinic crystal symmetry ($P2_1 /c$) \cite{Grube1999,Robinson2006}. Recently, continuous structural transitions at zero temperature started to attract interest as they can induce quantum critical lattice fluctuations \cite{Zacharias2015}. Indeed, experimental studies of the nonmagnetic analogue LaCu$_{6-x}$Au$_x$ suggested the possibility of an elastic quantum critical point of a monoclinic-orthorhombic transition without any magnetic transition \cite{Poudel2016}. Moreover, the structural and magnetic transitions in the rare-earth homologs CeCu$_{6-x}$Ag$_x$ and CeCu$_{6-x}$Pd$_x$ have been studied in detail \cite{Poudel2015}. While the magnetic QCP in CeCu$_{6-x}$Ag$_x$ occurs in the orthorhombic phase and is well separated from the structural phase transition, in CeCu$_{6-x}$Pd$_x$, the magnetic QCP occurs well within the monoclinic phase. Thus, CeCu$_{6-x}$Au$_x$ provides the unique setting to study the interplay between magnetic and elastic quantum critical fluctuations. 

In CeCu$_6$ the strong Kondo effect inhibits long-range antiferromagnetic order which, however, can be induced by applying negative chemical pressure. This is accomplished by replacing Cu by larger transition-metal ions, e.g., $M$~= Au \cite{HvL1994}, Ag \cite{Kuechler2004}, Pd, or Pt \cite{Sieck1996}. 
Here, we determine the critical Au concentrations of the magnetic and elastic zero-temperature transitions by tracking the antiferromagnetic and monoclinic phase boundaries down to $T \approx 30\,$mK as a function of both chemical and hydrostatic pressure. Employing neutron-diffraction, thermal-expansion, specific-heat, and magnetization measurements allows to compare the critical behavior of samples with varying distance to the possible elastic QCP and thereby to extract the impact of the elastic fluctuations on the magnetic behavior and vice versa. 

Single crystals of CeCu$_{6-x}$Au$_x$ with varying Au concentrations were grown by the Czochralski method under high-purity argon atmosphere. The gold content and occupancy of the Cu(2) site were determined by atomic absorption spectroscopy and single-crystal X-ray four-cycle diffraction analysis, respectively. We studied the structural phase transition of CeCu$_{6-x}$Au$_x$ samples with $x$~= 0.0, 0.101(1), 0.134(4), 0.150(1), and 0.155(3) by single-crystal elastic neutron scattering on the thermal triple-axis spectrometer 1T instrument at the Laboratoire L\'eon Brillouin, CEA Saclay, France. Measurements were performed using pyrolythic graphite as both monochromator and analyzer with a energy of $k_i=k_f=14.7\,$meV. The samples were mounted in a closed-cycle refrigerator with $2\,\text{K} \leq T \leq 300\,$K. The thermal expansion was measured with a homemade capacitance dilatometer and the heat capacity by using the semi-adiabatic heat-pulse technique. The magnetization measurements were performed in a commercial SQUID magnetometer (Quantum Design, San Diego, CA). 

\begin{figure}[t!]
\includegraphics{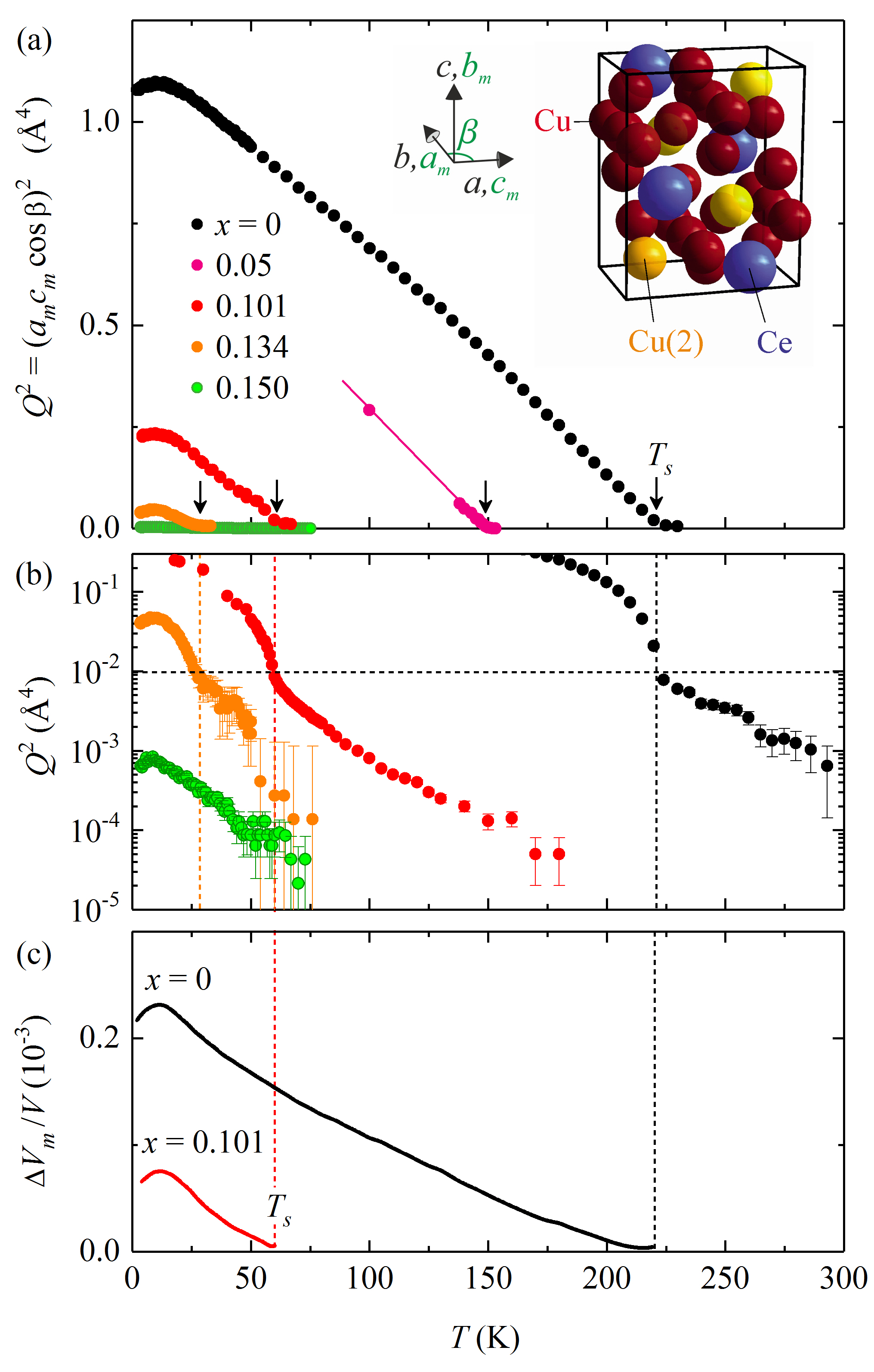}
\caption{\label{fig:mono} (Color online) (a) Monoclinic order parameter $Q$ of CeCu$_{6-x}$Au$_x$ for several $x$, $0 \leq x \leq 0.15$ as a function of temperature $T$. The inset shows the crystal structure. (b) Precursor of the monoclinic transition observed in $Q$. (c) Volume change $\Delta V_m/V$ due to the monoclinic distortion.}
\end{figure}

At room temperature, CeCu$_{6-x}M_x$ crystallizes in the orthorhombic CeCu$_6$-type structure. In this structure, five of the six Cu sites of the primitive unit cell are non-equivalent and, consequently, surrounded by different coordination polyhedra. The Cu(2) site has the largest volume and is occupied initially when Cu is being replaced by larger metal ions $M$ \cite{Ruck1993}. A tiny monoclinic distortion reduces the excess volume of the Cu(2) site $\Delta V_\text{Cu(2)}$ although the unit-cell volume of the monoclinic phase is larger than that of the orthorhombic phase. To distinguish between monoclinic and orthorhombic notations, which differ by a permutation of the axes, we use in the following the subscripts $m$ for the monoclinic notation [see inset of Fig.\,\ref{fig:mono}(a)]. The corresponding structural phase transition, which can be viewed as a small shear strain of the $a_m c_m$ planes, emerges from a softening of transverse acoustic phonons. Neutron-scattering experiments \cite{Noda1985}, thermal-expansion \cite{Grube1999}, ultrasound-velocity \cite{Goto1988,Finsterbusch1996}, and resistivity measurements \cite{Gratz1987} unambiguously reveal a continuous, second-order transition. Measurements under pressure indicate that this transition is extremely sensitive to volume changes and can easily be suppressed by reducing $\Delta V_\text{Cu(2)}$, either by external hydrostatic pressure \cite{Grube1999,Oomi1988}, by replacing Ce with smaller rare-earth ions $R$ \cite{Yamada1987,Vrtis1990,Nakazato1990a}, or by substituting larger metal ions for Cu(2) \cite{Grube1999}. The CeCu$_6$ structure is homogeneously maintained in CeCu$_{6-x}$Au$_x$ for $x \leq 1$ \cite{Mock1994}.

Below the structural transition temperature $T_s$ the monoclinic distortion leads to a twinning with $\left\{ 100 \right\}_m$ as common twin plane. The order parameter $Q$ is given by the monoclinic strain $a_m c_m \cos\beta$. Using elastic neutron scattering, we determined the temperature dependence of $Q$ by measuring the lattice parameters and the splitting angle of the Bragg peak $(200)_m$, which is twice the deviation angle from $90^\circ$. In Fig.~\ref{fig:mono}(a), $Q^2$ is plotted as a function of $T$. In accordance with the expected mean-field behavior $Q \propto (T_s-T)^{1/2}$, $Q^2$ shows a roughly linear $T$ dependence over a wide temperature range as previously observed for CeCu$_6$ \cite{Nakazato1990a}. Linear extrapolations of $Q^2(T)$ to $Q^2 = 0$ are used to determine $T_{s}$. With increasing $x$, $T_{s}$ shifts to lower $T$. For $x\geq 0.150$ no long-range monoclinic distortion could be detected. 

Towards low temperatures in the monoclinic phase, the measurements of $Q^2$ exhibit a maximum and an unusual downturn. A similar, albeit stronger, behavior has been reported for SmCu$_6$ where it was attributed to a negative valence change of the Sm ion \cite{Nakazato1990a}. The maximum of $Q^2(T)$ in CeCu$_{6-x}$Au$_x$ occurring around 10\,K maybe attributed to the Kondo effect which entails virtual transitions of the 4$f$ electron to the Fermi level, thus minimally decreasing the effective 4$f$ occupancy of Ce leading to an effective reduction of the Ce-ion size. In line with this argument, the temperature of the $Q^2$ maximum  corresponds to the (single-ion) Kondo temperature of $T_K \approx 6\,$K \cite{Schlager1993}. By employing measurements of the thermal expansion and the monoclinic distortion angle $\beta$ we estimate the volume change generated by the monoclinic distortion $\Delta V_m\approx V(x)-V(x=0.15)$ with the volume above $T_s$ subtracted [see Fig.\ref{fig:mono}(c)]. The temperature dependence of $\Delta V_m$ closely resembles that of $Q^2$ with an even more significant reduction below $T_K$. 

Above $T_s$, the structural transitions are accompanied by precursors extending to $\approx 2T_s$ [see Fig.\,\ref{fig:mono}(b)]. As the precursors appear in both Au-alloyed and stoichiometric CeCu$_6$, they are unlikely to arise from sample inhomogeneities but rather point to critical fluctuations of the order parameter, as expected for continuous phase transitions. 
The crossover from a finite $Q$ for $T < T_s$ to fluctuations occurs at the same $Q_{exp}$, independent of $x$ [horizontal dashed line in Fig.\ref{fig:mono}(b)]. Below $Q_{exp}$, the monoclinic splitting of the $(200)_m$ peak can no longer be resolved as two separate peaks because of the finite experimental energy resolution. Here, $Q^2$ is inferred from the observed broadening beyond the experimental resolution due to fluctuations within the time window of our neutron scattering measurements ($\approx 10^{-12}$\,sec). The fact that the data of $Q^2$ for $x=0.150$ fall short of $Q_{exp}$ by a factor of $10$ at the lowest temperature suggests that $x=0.150$ is a little larger than the critical concentration for $T_s=0$.  

\begin{figure}[t!]
\includegraphics{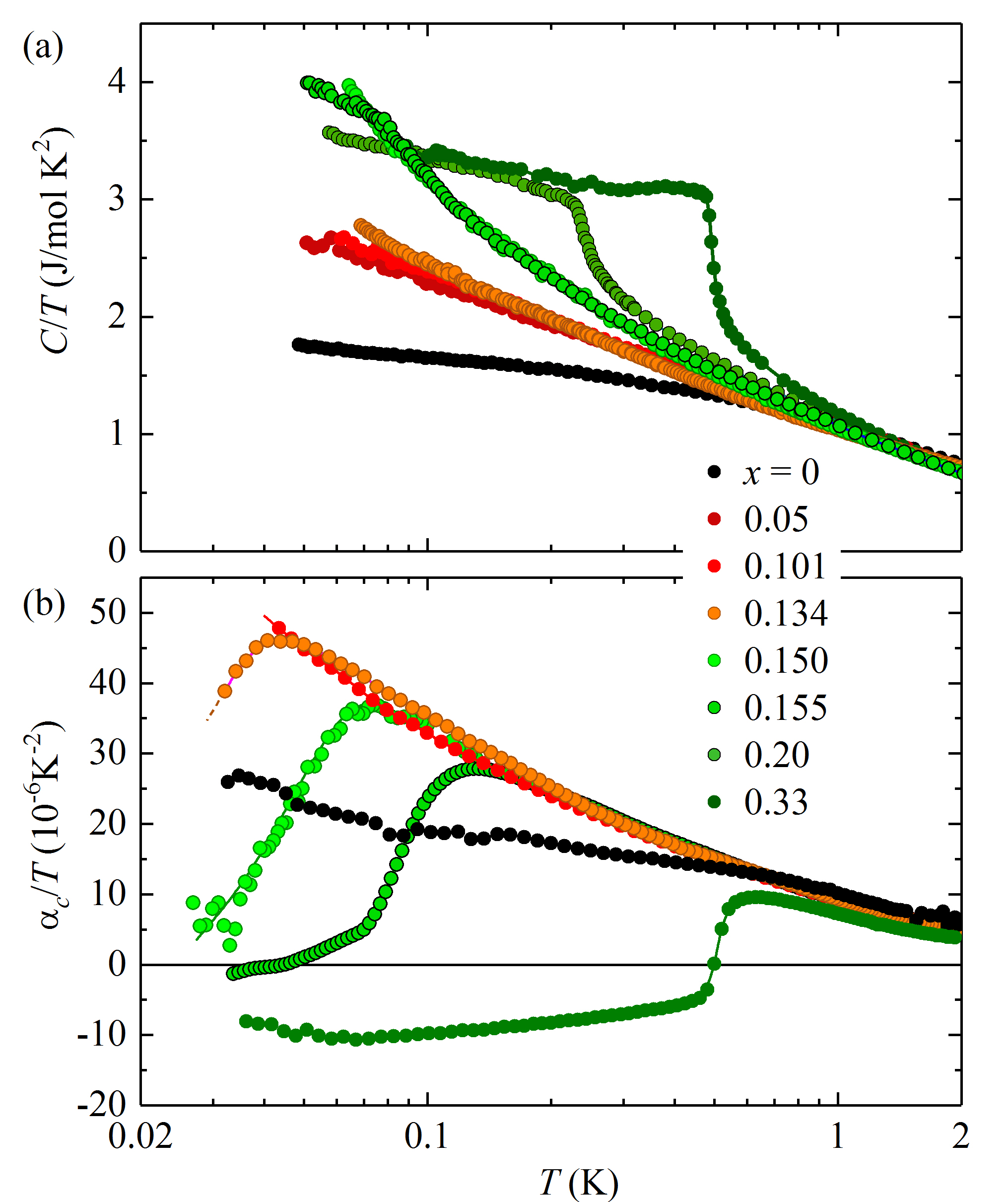} 
\caption{\label{fig:alpha} (Color online) (a) The specific-heat coefficient $C/T$ and (b) the linear thermal-expansion coefficient $\alpha_c$ divided by temperature $T$ as a function of $T$. The data of the $x=0.150$ sample were measured along the $a$ axis and for comparison multiplied by constant factor of $-3.5$.}
\end{figure}

We now turn to the magnetic phase transition. As a consequence of its second-order nature, the transition anomaly observed in the specific heat $C$ disappears when $T_N$ approaches zero \cite{HvL1998}. At a pressure-induced QCP, on the other hand, the anomalies of the linear thermal expansion coefficients $\alpha_i$ along the $i=a,b,c$ axes remain very large. 
This results from the Ehrenfest relation between the changes of $\alpha_i$ and $C$ at $T_N$ and the uniaxial pressure dependence of $T_N$: $\Delta \alpha_i = \Delta C(VT_N)^{-1}(\partial T_N /\partial \sigma_i)$, with $i=a$, $b$, $c$.
The normalized ratio $T_N^{-1} \partial T_N/\partial \sigma_i$ is synonymous with the Gr\"uneisen ratio $\Gamma_i$ of the antiferromagnetic phase with the ordering temperature as characteristic energy scale $E^*$. As a hallmark of QCPs, the Gr\"uneisen ratio diverges by approaching the QCP because $E^* \rightarrow 0$ \cite{Zhu2003}. This leads to an anomaly of $\alpha_i$ enhanced by $\Gamma_i$  compared to that of $C$. To resolve the magnetic transition at very low temperatures we therefore recorded $\alpha_i$ as a function of $T$ along the principal orthorhombic crystallographic axes $i$~= $a$, $b$, $c$, neglecting the small monoclinic distortion. The specific-heat and thermal-expansion results are displayed as $C/T$ and $\alpha_c/T$ in Fig.~\ref{fig:alpha}(a) and (b), respectively. For $x \geq 0.134$ the transitions are clearly visible in $\alpha_i$ and exhibit the expected increase of $T_N$ towards higher gold concentrations $x$. $\alpha_c(T)$ of the $x=0.134$ single crystal reveals at very low temperatures a downturn, indicative of a phase-transition onset. The base temperature of our experiment of $\approx 30\,$mK, however, prevents us from observing a full transition. The positive deviation from the $C/T ~ \ln T$ dependence for $x = 0.134$ below $\approx 100\,$mK supports the existence of a finite-$T$ magnetic transition (cf. $C/T$ vs. $T$ for $x = 0.150$).

\begin{figure}[t!]
\includegraphics{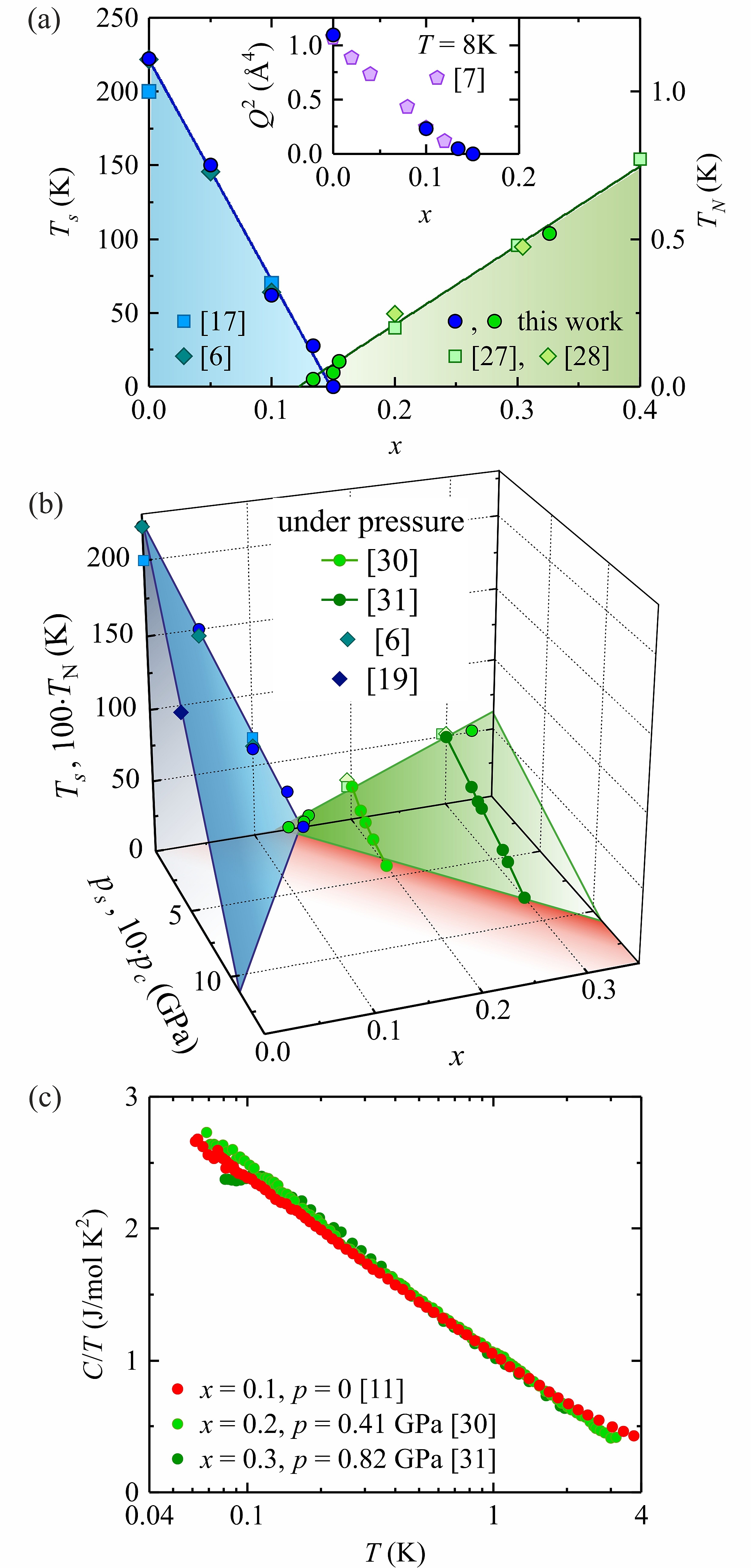}
\caption{\label{fig:PD} (Color online) (a) ($x$,$T$)-phase diagram of CeCuAu$_{6-x}$Au$_x$. The inset shows the decrease of the order parameter $Q$ with $x$.  
(b) The phase diagram extended by the hydrostatic pressure $p$ as additional control parameter. The $T_s(x=0)$ value under pressure is estimated by using the $p$ dependence of $T_s$ from Ref.\,\cite{Oomi1988}. (c) The specific heat coefficient $C/T$ of crystals that have been tuned to the magnetic quantum-critical phase boundary by varying $x$ and $p$. The identical scaling behavior has been reported for CeCu$_{5.95}$Pd$_{0.05}$ \cite{Sieck1996} and CeCu$_{5.8}$Ag$_{0.2}$ \cite{Kuechler2004}.}
\end{figure}

We show in Figure~\ref{fig:PD}(a) the results of $T_N(x)$ and $T_s(x)$ together with literature data \cite{Finsterbusch1996,Grube1999,Okumura1998,Pietrus1995}. $T_N$ was taken as the temperature of the midpoint of the transition in $\alpha_c(T)$ shown in Fig.~\ref{fig:alpha}(b). For $x \lesssim 0.3$, $T_N$ roughly follows the linear $x$ dependence observed for higher $x$. In particular, the data do not exhibit a sublinear decrease towards $T_N \rightarrow 0$ expected for the crossover to a three-dimensional antiferromagnetic quantum criticality \cite{Garst2008}. The data, therefore, confirm the two-dimensional character of the fluctuations in the investigated $T$ range.

We note a tiny overlap between the two different types of order. This is experimentally substantiated for the sample x = 0.134 which shows a clear onset of a magnetic transition at $\approx 0.04$\,K [see Fig.~\ref{fig:alpha}(b)] and a clear signature of the structural transition at $T_s = 27.5$\,K. This is a very strong hint that the two transitions are separate and do not meet at a quantum (multicritical) point at $T = 0$ as suggested by Robinson \textit{et al.} \cite{Robinson2006}. Looking at the ensemble of samples, $T_N$ vanishes at $0.1 \leq x < 0.134$, and $T_s$ vanishes for $x > 0.150$. The fact that the magnetic and monoclinic phases in CeCu$_{6-x}$Au$_x$ coexist in a tiny concentration overlap is independent of the exact $x$ dependencies of $T_N(x)$ and $T_s(x)$. This behavior is to be contrasted with that of two extreme cases: CeCu$_{6-x}$Ag$_x$ where the monoclinic distortion ends for Ag concentrations well below the onset of magnetic order, and CeCu$_{6-x}$Pd$_x$ where $T_s$ does not depend at all on the Pd content \cite{Poudel2015}.

The proximity of the magnetic and elastic QCP in CeCu$_{6-x}$Au$_x$ is accidental. This is underpinned by a comparison of the reported pressure dependences of the magnetic and monoclinic transitions at various Au concentrations shown in Fig.\,\ref{fig:PD}(b): the hydrostatic pressure dependences of $T_s$ and $T_N$ are both negative \cite{Grube1999,Oomi1988,Stockert2002,Bogenberger1995}, thus separating the two QCPs with increasing pressure. On the other hand, the character of the magnetic quantum phase transition is preserved. Figure\,\ref{fig:PD}(c) shows the specific heat $C/T$ on a logarithmic $T$ scale for $x = 0.1$ at $p = 0$ \cite{HvL1994}, $x = 0.2$ at $p = 4.1\,$kbar \cite{Stockert2002} and for $x = 0.3$ for $p = 8.2$\,kbar \cite{Bogenberger1995}. 
Not only is the scaling behavior $C/T \propto \ln(T_0/T)$ identical, but also the characteristic temperature $T_0$ governing the slope of the logarithmic divergence is preserved. $T_0$ is analogous to the amplitude of the divergence for $T > T_c$ and $T < T_c$ in classical second-order phase transitions at finite $T_c$. Moreover, even the absolute values of the anomaly of the specific heat and hence the entropy at any temperature are identical, indicating that along the boundary of magnetic vs. nonmagnetic ground states always the same number of degrees of freedom is released. Thus an accidental degeneracy of $T_s$ and $T_N$ at a concentration at $x=0.14$ is not expected to induce a change in the unusual magnetic quantum-critical behavior.

The insensitivity of the magnetic QCP to the monoclinic distortion might arise from the fact that the critical elastic and magnetic fluctuations evolve in two different crystallographic planes which are orthogonal, and therefore independent of each other \cite{Grube2017}. 
On more general grounds, the coupling of elastic degrees of freedom and order parameters that couple bilinearly to strain fluctuations was analyzed theoretically \cite{Una2016,Paul2017} with the result that long-range strain fields cause such transitions to be mean-field like. This is clearly not what is seen in case of the magnetic transition, demonstrating that the unconventional quantum criticality couples only weakly to strain fluctuations. This is further corroborated by our magnetization measurements (not shown) that do not reveal any change in the magnetization across the structural transition, confirming that the relevant magnetic fluctuations do not couple to the structural transition. 

We have shown that the combination of chemical substitution and hydrostatic pressure allows to separate the magnetic from the structural transition.
Furthermore, elastic neutron scattering clearly detected critical fluctuations above a structural QPT. 
The comparison between crystals with different $x$ that are driven by hydrostatic pressure to the magnetic QCP and have different distances to the structural instability, demonstrates that the sensitive balance of Kondo effect and RKKY interaction is hardly affected by the monoclinic distortion and that the structural fluctuations do not relate to the unconventional quantum critical behavior. In view of the strong electron-lattice coupling of heavy fermion systems in general \cite{Luthi1987}, this finding at first sight is surprising. The universality of the unusual magnetic quantum criticality thus appears to be surprisingly robust and independent of the proximity to the structural transition.

We thank M. Garst, P. Gegenwart, and J. Schmalian for helpful discussions, and the Deutsche Forschungsgemeinschaft for support in the frame of Research Unit 960 "Quantum Phase Transitions".

%

\end{document}